\documentclass[sigconf]{acmart}
\usepackage{graphicx}
\AtBeginDocument{%
  }

\copyrightyear{2024}
\acmYear{2024}
\setcopyright{acmlicensed}\acmConference[JCDL '24]{The 2024 ACM/IEEE Joint
Conference on Digital Libraries}{December 16--20, 2024}{Hong Kong, China}
\acmBooktitle{The 2024 ACM/IEEE Joint Conference on Digital Libraries (JCDL
'24), December 16--20, 2024, Hong Kong, China}
\acmDOI{10.1145/3677389.3702537}
\acmISBN{979-8-4007-1093-3/24/12}

\acmConference[JCDL '24]{The 2024 ACM/IEEE Joint
Conference on Digital Libraries}{December 16--20, 2024}{Hong Kong, China}
\acmISBN{979-8-4007-1093-3/24/12}

\begin{document}

\title{Collaborative Data Behaviors in Digital Humanities Research Teams}

\author{Wenqi Li}
\email{wenqili@pku.edu.cn}
\affiliation{
\institution{Department of Information Management,}
\country{Peking University, Beijing, China}
}
\author{Zhenyi Tang}
\email{tangzhenyi@pku.edu.cn}
\affiliation{
\institution{Department of Information Management,}
\country{Peking University, Beijing, China}
}
\author{Pengyi Zhang}
\email{pengyi@pku.edu.cn}
\affiliation{
\institution{Department of Information Management,} 
\country{Peking University, Beijing, China}
}
\author{Jun Wang}
\email{junwang@pku.edu.cn}
\affiliation{
\institution{Department of Information Management,} 
\country{Peking University, Beijing, China}
}
\begin{abstract}
The development of digital humanities necessitates scholars to adopt more data-intensive methods and engage in multidisciplinary collaborations. Understanding their collaborative data behaviors becomes essential for providing more curated data, tailored tools, and a collaborative research environment. This study explores how interdisciplinary researchers collaborate on data activities by conducting focus group interviews with 19 digital humanities research groups. Through inductive coding, the study identified seven primary and supportive data activities and found that different collaborative modes are adopted in various data activities. The collaborative modes include humanities-driven, technically-driven, and balanced, depending on how team members naturally adjusted their responsibilities based on their expertise. These findings establish a preliminary framework for examining collaborative data behavior and interdisciplinary collaboration in digital humanities.
\end{abstract}


\maketitle

\section{Introduction}
Digital humanities (DH) have flourished over the last few decades, continually expanding the variety of data and technologies used in the humanities studies. The scholarly activities around data and collaboration are becoming more essential to humanities scholars as they adopt more data-intensive approaches and engage into more multidisciplinary teams \cite{pacheco_digital_2022, given_information_2018}. As the field continues to grow, DH researchers are seeking more curated data and collections, tailored data tools and services, and a highly collaborative research environment. A comprehensive understanding of their collaborative data behaviors is essential to promote the interdisciplinary collaboration and inform the design of the research infrastructure in DH field.

Data behavior studies focuses on the observable actions and reactions of users when they encounter, seek, create, or use data for individual or collaborative tasks \cite{zhang2023conceptualizing}. Existing works have examined humanities scholars' data behaviors, such as seeking and sharing ~\cite{li_analysing_2024, late2024share}, as well as data activities and workflows in DH research ~\cite{ma_data_2020, hoekstra_data_2019,late_interacting_2022}. However, the collaborative dynamics around these data activities in DH research teams remain unexplored.

Collaborative data behavior spans a wide range of activities, such as creating or analyzing data in a team, sharing it with others, or reusing others' data ~\cite{koesten2019collaborative}. Although collaborative data behavior has not been investigated in depth, a lot could be drawn from collaborative information behavior studies ~\cite{karunakaran2013toward}, such as the collaborative process, driving factors, and division of labour ~\cite{tao2017collaborators,Jin2015Collaborative,YAN2018Construction,foley2010division}.

In the field of DH, more attention needs to paid to collaborative data behavior as the research nature shifts from an individual-centric and hermeneutic approach to a more collaborative and data-driven approach ~\cite{given2015collaboration}. Understanding the collaborative data behaviors in DH teams is essential for optimizing their data workflow, ensuring data quality, thus encouraging and improving interdisciplinary collaboration. However, there's a lack of research focused on the collaborative data behaviors in DH teams, despite some studies examining the benefits and challenges of DH collaboration, collaboration tools, and team development ~\cite{siemens_its_2009,siemens_tale_2011,siemens_time_2010,poole2017greatly}.

Therefore, we aim to explore the collaborative data behaviors in DH teams to address this gap. In this paper, we present preliminary findings on the research question: How do interdisciplinary researchers collaborate on data activities within digital humanities research teams?

\section{Related Works}
\subsection{Data Practices in Digital Humanities}
The advent of DH is transforming the humanities research towards a more data-intensive approach ~\cite{pacheco_digital_2022}, thus raising scholars' attention to humanities scholars' data behaviors and data practices in DH. For example, Borgman examined data scholarship in the humanities with detailed case studies, establishing a preliminary understanding of how humanities scholars work with data ~\cite{borgman_big_2015}. Li et al. identified characteristics of humanities scholars' data seeking behaviors and their variations across research approaches and seeking phases ~\cite{li_analysing_2024}.

A few studies explored the types and processes of data practices in digital history. Ma and Xiao summarized a comprehensive data workflow \cite{ma_data_2020}. Hoekstra and Koolen proposed the concept of ``data scopes'' to demonstrate types of data transformation activities in history research, including selection, modeling, normalization, classification, and linking \cite{hoekstra_data_2019}. Late and Kumpulainen identified various data activities, such as acquiring access to data, analyzing metadata, annotating, and visualizing data \cite{late_interacting_2022}. Yet few research explored DH researchers' collaboration in these data activities.

\subsection{Collaborations in Digital Humanities}
Recent studies indicate that humanities research is increasingly team-based and collaborative \cite{given_information_2018}. In particular, DH researchers often deal with large text corpora that must be digitized before analysis, leading to collaboration within large research teams \cite{given2015collaboration}. Collaboration facilitates the exchange of insights and completion of tasks when multiple researchers work towards a collectively shared goal \cite{sonnenwald_scientific_2007}. Through collaboration, DH researchers can leverage expertise, share costs and resources, access new tools, and establish standards and best practices \cite{harvey_digital_2010}.

However, collaborations in DH face challenges, such as developing trust and understanding, achieving consensus on language and terminology, equal contributions, accountability and responsibility \cite{poole2017greatly}. Additionally, specific coordination activities such as meeting scheduling, task allocation, information sharing, and training are necessary \cite{siemens_time_2010}. The roles of ``hybrid people'' or ``translators'' can help identify and resolve conflicts, harmonize terminology across disciplines, and facilitate interdisciplinary collaboration \cite{edmond_collaboration_2015,siemens_tale_2011}. Existing studies have identified different strategies for division of labor in collaborative information seeking or collaborative writing, such as parallel, sequential, and reciprocal ~\cite{foley2010division, sharples1993adding}. However, there's little research exploring the division of labor or collaborative patterns in DH research teams, except for a few recent studies that revealed collaborative roles in specific DH projects \cite{oberbichler2022integrated, li_curating_2024}.

\section{Methods}
We employed focus group interviews for data collection, through which we were able to obtain a wide range of perspectives from different team members within a single session~\cite{gorman2005qualitative}. The focus group interviews were conducted during two digital humanities summer courses held in 2022 (Course A) and 2023 (Course B), both integrated lectures with team projects and spanned two weeks.  

\begin{figure}[h]
  \centering
  \includegraphics[width=\linewidth]{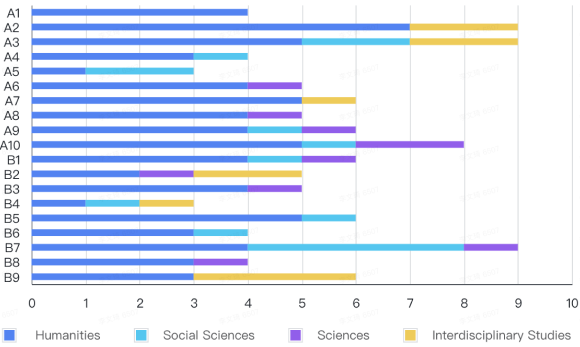}
  \caption{Academic backgrounds of focus group members}
  \label{fig:participants}
  \Description{A stacked bar chart showing numbers of participants from humanities, social science, science and engineering, and interdisciplinary backgrounds in each team.}
\end{figure}

Research teams voluntarily signed up for focused group interviews. In Course A, we recruited 10 research teams, totaling 56 individuals. In Course B, we recruited 9 research teams, comprising 40 individuals. Most teams consisted of 3-6 members, with four having 8-9 members. Most participants were graduate students majored in humanities. According to their registration information, participants with humanities backgrounds generally had limited technical skills. Figure \ref{fig:participants} details the team members' academic backgrounds. 

In the final stage of each course, as most teams were wrapping up their research projects, we conducted focus group interviews with the recruited teams. We probed into the data activities involved in their team projects, their expectations and workflows, and the challenges encountered. We particularly focused on team collaboration, including task division and individual roles. Four researchers independently facilitated the focus groups, either online or in person accommodating participants' preferences. Each focus group session lasted 40-50 minutes, with the facilitator ensuring active participation from all members. The sessions were recorded and transcribed using voice transcription software.

We conducted inductive coding on interview transcripts. Five researchers participated in coding and discussions. Initially, two coders independently coded the transcripts: one coded transcripts from Course A and another coded Course B. The research team held regular discussions to refine the coding scheme. Then, the third coder coded all the transcripts using the finalized coding scheme, which included three code categories: primary data activities, supportive data activities, and collaborative modes. We note that this is only the preliminary coding scheme at this exploratory stage.


\section{Findings}
Our coding results show two categories of data activities in DH research. \textit{Primary data activities} are those that directly contribute to the core research. They include data selection and collection, data modeling, data processing, data analysis, and data interpretation. These activities are integral to the development and execution of the research, forming the foundation of research findings. \textit{Supportive data activities} support the primary data activities. They include orientation, data quality assurance, and interdisciplinary communication. These activities facilitate the smooth operation and effectiveness of the primary data activities.

We further found that participants naturally adjusted responsibilities across data activities based on technical and humanities expertise. Three collaborative modes emerged from coding: humanities-driven, technically-driven, and balanced. In \textit{humanities-driven collaborations}, humanities participants drive the direction while technical participants provide supports. In \textit{technically-driven collaboration}, participants with technical skills set the goal and standard, and humanities participants contribute their domain expertise as needed. The \textit{balanced collaboration} mode involves an equal partnership, where technical and humanities participants integrate their expertise for comprehensive and negotiated outcomes. Table \ref{tab:modemapping} presents varying collaborative modes across data activities. 

\begin{table}[]
\caption{Collaborative modes across different data activities}
\label{tab:modemapping}
\resizebox{\columnwidth}{!}{%
\begin{tabular}{lll}
\hline
\textbf{Data Activity Category} & \textbf{Data Activities}        & \textbf{Collaborative Modes} \\ \hline
Primary data activities  & Data selection \& collection & Humanities-driven \\  & Data modelling   & Humanities-driven \\ & Data processing  & Technically-driven\\  & Data analysis \& interpretation & Balanced \\ 
Supportive data activities & Orientation & Balanced\\  & Data quality assurance  & Balanced\\ & Interdisciplinary communication & Balanced  \\ \hline
\end{tabular}%
}
\end{table}

\subsection{Primary Data Activities}
\subsubsection{Data Selection and Collection}
Data selection and collection are mostly driven by humanities participants, who tend to be more aware of the availability and usefulness of related sources because of their familiarity with the topic. In most focus groups, humanities participants select data sources based on their domain relevance and quality, including authority and accuracy. They would determine the starting point and ideal scope of their data collection. Subsequently, participants with technical backgrounds would evaluate the feasibility to access or process the data and make suggestions of available digital versions accordingly. 

For instance, a participant in Group A8 mentioned that the book ``Water Margin'' has several versions, including the seventy-chapter version, the hundred-chapter version, and other annotated versions, most of which may contain errors. A participant majored in Chinese initially selected a well-recognized version from a reputable publisher. However, access to this edition was limited to PDF format, which posed challenges for text conversion necessary for subsequent computational analysis, as suggested by a participant with technical expertise. After thorough discussion and collaboratively searching for available text versions, humanities participants made final decisions on an accessible text version, which was not as authoritative but more suitable for machine processing.

\subsubsection{Data Modelling}
Data modeling refers to the activity of defining and structuring data elements and their relationships to support effective data processing and analysis. Many focus group participants emphasized the importance of humanities scholars in driving data modeling, which directly determines the usefulness of data for the DH research. Humanities scholars contribute a research question-oriented perspective, ensuring the appropriate granularity and comprehensiveness of the data structure.

Several humanities participants noted that the lack of early involvement of humanities scholars in the data modeling process caused the data less useful. For example, a participant from Group A3 mentioned that their team set a goal to build a knowledge graph of paintings from the Han Dynasties, but few art history scholars were involved in the initial data modeling process. Although technical participants offered a new approach for them to decompose the characteristics of the paintings, they were unsure of the purpose, questioning: \textit{``what research questions can be addressed afterward?''}

\subsubsection{Data Processing}
Data processing in most focus groups was driven by participants with technical backgrounds, except for a few groups like A1 that lacked technical members. DH research relies on large quantities of machine-operable data for computational analysis. Therefore, technically skilled participants, familiar with data formats and processing pipelines, often drive the workflow. However, this driving role might not be purposefully assigned, but rather naturally emerged. As a participant in Group B3 mentioned they didn't have clearly defined roles at the beginning of data processing: \textit{``I initially suggested that the three of us with humanities backgrounds could manually extract the structured data. But then our technical member offered writing codes to automate the process, which greatly increased our efficiency.''}

Once the data processing goal and workflow are confirmed, humanities participants contribute in further annotation and proofreading. Many focus groups mentioned that the accuracy of data pre-processing and automated extraction is not as satisfying. Much of semantic extraction work relies on the humanities’ interpretation and judgment. A humanities scholar in Group A1 mentioned that he was the only one with the domain knowledge to extract the place names, because the scholarly discussions and determination around these ancient place names are constantly evolving.

Another important aspect of data processing is the documentation and handling of intermediate products, which is also mostly led by technical participants. For instance, Group A6 mentioned that they maintained data summary tables and task logs for every step of data cleaning, in case \textit{``anyone needs to trace our work later or seek evidence to support a conclusion.''}

\subsubsection{Data Analysis and Interpretation}
Data analysis and interpretation needs balanced contributions from both humanities and technical sides. Some participants mentioned that this is a process of ``mutual inspiration'' and ``continuous iteration'', where technically skilled participants provide initial analysis results and visualizations through methods like social network analysis, topic analysis, and GIS mapping. Humanities participants then attempt to interpret these results and propose ways to delve deeper or optimize the analysis and presentation. This iterative process may lead to unexpected findings and more streamlined research approaches. For example, a participant with technical background in Group B2 stated that he almost gave up on his topic modeling results as he couldn't see any useful difference, but \textit{``thanks to our humanities teammate, who immediately spotted a stable pattern. Otherwise, our efforts would have been a dead end.''} His humanities teammate then added: \textit{``Our ways of thinking are quite different. My science teammates pulled me out of the chaos of my humanities mindset, showing me that I can work step-by-step to achieve results without isolating myself and forcefully creating narratives like I used to do.''}

\subsection{Supportive Data Activities}
All supportive data activities need balanced contribution from both humanities and technical members.
\subsubsection{Orientation}
Once a DH team is formed, onboarding members is important for aligning project goals and familiarizing them with tools and procedures. For example, participants in Group A6 mentioned that at the beginning, humanities participants familiarized the technical participants with relevant humanities databases. Technical participants, on the other hand, introduced the software they needed to use to the humanities participants. Thus everyone understood the capabilities of the database and tools and had a clearer vision of project direction.

Moreover, some participants emphasized that all team members, particularly the later participants, should learn about the overall data workflow cohesively. A later participant in Group A3 mentioned that insufficient orientation diminished his motivation: \textit{``I was simply given an Excel sheet to fill out without even knowing the purpose of my tasks. I felt a bit lost, like I was only seeing the tail of an elephant without knowing what the whole elephant looked like.''}

\subsubsection{Data quality assurance}
This supportive activity is mostly integrated into the data processing. It includes establishing data processing standards and quality assurance mechanisms, data accuracy checks, and more. Since lots of uncertainties exist in the transformation of humanities data, a pre-defined standard or rules of data processing is crucial, especially for larger data processing teams. Both technical and humanities participants need to discuss the standard and ensure everyone who participated in the data processing can follow. For example, the group leader in Group A2 mentioned that she wrote a processing procedure document that everyone could refer to as needed. 

Even with standards in place, there remain quality issues, particularly with manual annotation, which manifests in three aspects: inconsistent annotation formats that affect machine-operability, domain-related uncertainties or academic disagreements, and incorrect or missed annotations. Participants suggested that both technical and humanities participants should work together towards a feasible quality assurance mechanism that balances the data accuracy and cost of human resources. When performing data quality checks, most teams had technical participants verifying data formats and humanities participants reviewing content accuracy.

\subsubsection{Interdisciplinary communication}
Effective communication between technical and humanities participants, despite their different roles and responsibilities, is considered fundamental to ensure the project feasibility and usefulness of data. For example, the leader of Group A8 mentioned that whenever a key data issue needs to be resolved, they consult both humanities and technical participants and discuss based on their collective insights. Participants in Group B3 also noted: \textit{``Turning a historical problem into a data problem is an abstraction process that requires both coding skills and historical knowledge. In our group, every abstraction task involves communication between at least one technical member and one historian to determine exactly what is needed."}

Different ways of thinking and specialized knowledge can make communication challenging. Several focus group mentioned that social science or interdisciplinary team members frequently serve as "bridges" to facilitate communication. Many participants suggested their successful teamwork benefited from avoiding the use of jargon and clear communication of terminology to ensure mutual comprehension. For example, a participant in Group A6 stated: \textit{``We used formalized language to clarify our needs to technical teammates, avoiding broad requests like displaying all relationships from a text input. Likewise, she explained feedback in terms we could understand instead of directly copying Python codes.''}

On the other hand, the different ways of thinking and communication can also inspire each other. For example, a humanities participant in Group A10 reflected on the complementary advantages of working across disciplines:
\textit{``I proposed an abstract concept that I haven't quite figured out myself. The next day, I was surprised to see our technical teammate had described this concept clearly and mathematically, using formalized equations.''}

\section{Conclusions}
In this paper, we provide preliminary insights into the collaborative data behaviors within interdisciplinary DH research teams. Through inductive coding, we identified seven types of data activities, categorized into primary and supportive data activities. We also found DH teams employed different collaborative modes in various data activities: humanities-driven, technically-driven, and balanced, depending on how team members contribute their expertise. These collaborative modes, observed in our data, have limitations as participants were students newly exploring digital humanities. While experienced digital humanities scholars who are proficient in both humanities and technology might collaborate in more integrated ways, our findings are particularly relevant for scholars who are beginning to explore digital humanities collaboration, offering insights that can help them quickly adapt and collaborate effectively. As they accumulate more experiences and skills, they may achieve deeper integration rather than the divisions observed in this study.

The identification of data activities and collaboration modes establishes a preliminary framework for further investigation of collaborative data behaviors, which is a new perspective for examining interdisciplinary collaboration in DH. Existing studies have developed collaborative process models revealing various collaborative activities and their influencing factors ~\cite{chung2016understanding,tabak2012non,robson2013building}. Our research, however, delved into the specific data activities to examine the collaborative mode within each activity, based on the varying skill requirements for collaborators in each activity. Comparing to DH team collaboration studies that analyze research roles and coordination \cite{siemens_its_2009, oberbichler2022integrated}, collaborative data behavior allows us to focus on more specific data activities that are core to DH research. Further investigation is needed to reveal more nuanced collaborative patterns, such as synchronous or asynchronous ~\cite{Zhang2022Research}, to achieve more seamless collaboration and transition. In-depth analysis of collaborative data behaviors can provide insights that directly enhance data quality and data work efficiency in digital humanities research, including the structuring of interdisciplinary DH teams, the curation of humanities data and digital collections, design implications for collaborative data handling tools, and more. Such improvements are crucial for facilitating collaborative DH research.

We acknowledge that the work is still at a very early stage. In future work, we will further investigate the nuances and dynamics of collaborative data behaviors within DH teams, including the role adjustments, strategies for division of labour, knowledge sharing, conflict resolution, and driving factors.

\begin{acks}
The study was supported by Chinese Library Association Research Grant 2022LSCKYXM-ZZ-YB002 and NSFC Grant No. 72010107003. We thank the anonymous reviewers for their insightful feedback.
\end{acks}

\bibliographystyle{ACM-Reference-Format}
\bibliography{sample-base}

\appendix

\end{document}